\documentclass{appolb}
\usepackage{epsfig}
\usepackage{amssymb}
\usepackage{amsmath}
\usepackage{placeins}
\usepackage[utf8]{inputenc}
\usepackage[polish,english]{babel}
\usepackage[OT4]{fontenc}
\usepackage[MeX]{polski}
\usepackage{float}
\usepackage{tikz}

\begin{document}
\title{Mean pion multiplicities in Ar+Sc collisions%
\thanks{Presented at CPOD 2016}
}

\author{Michał Naskręt
\address{University of Wroclaw, NA61/SHINE}
}

\maketitle

\begin{abstract}
Preliminary results for mean negatively charged pion multiplicities $\langle \pi^- \rangle$ using the $h^-$ method are presented for central Ar+Sc collisions at 13, 19, 30, 40, 75 and 150\textit{A} GeV/c beam momentum. The data were recorded by the NA61/SHINE detector at the CERN SPS. Starting with rapidity distributions ${dn}/{dy}$ the procedure of obtaining total multiplicities is presented. The mean number of wounded nucleons $\langle W\rangle$ extracted from the Glissando MC model is used to calculate the ratio $\langle \pi^- \rangle/\langle W\rangle$. The results are compared to those from other experiments and their dependence on colliding systems and collision energy is discussed.
\end{abstract}


\section{Introduction}
NA61/SHINE is a large acceptance fixed target experiment which studies final hadronic states in interactions between various particles and nuclei~\cite{na61det}. Pions are by far the most abundant particles produced in heavy ion collisions. Using the $h^-$ method one can select negatively charged pions out of all the produced particles~\cite{antoni}.

Central Ar+Sc collision measurements are very important in the physics programme of the NA61/SHINE experiment. Analysis of the data will help understand the physics of the onset of deconfinement and the critical point. Among the many different hadrons produced in high energy collisions, pions are the lightest and the most numerous. Thus, data on pion production properties is crucial for constraining the basic properties of models of strong interactions.

\section{$\pi^-$ rapidity distributions}
\begin{figure}
\centering
  \includegraphics[width=0.3\textwidth]{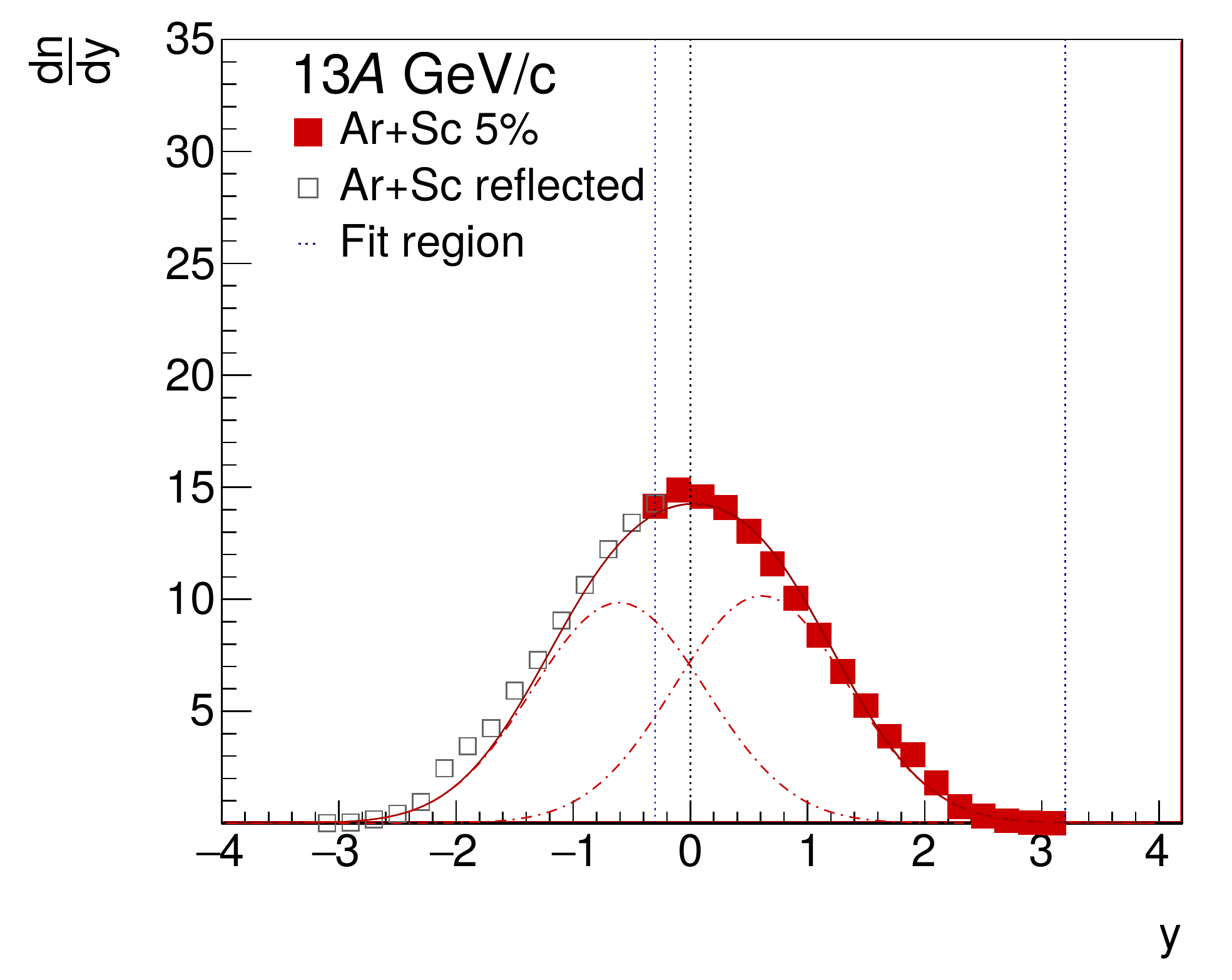}
  \includegraphics[width=0.3\textwidth]{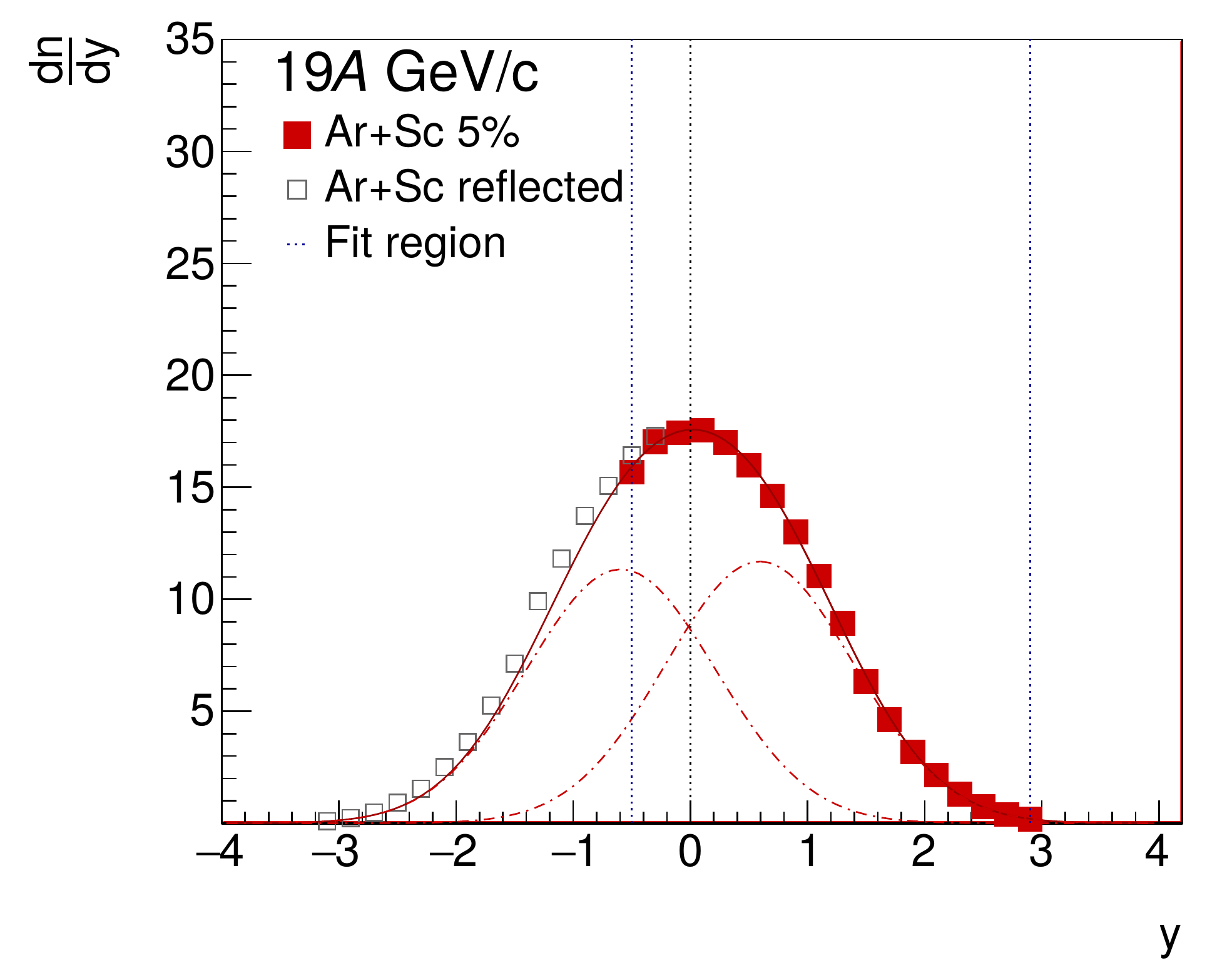}
  \includegraphics[width=0.3\textwidth]{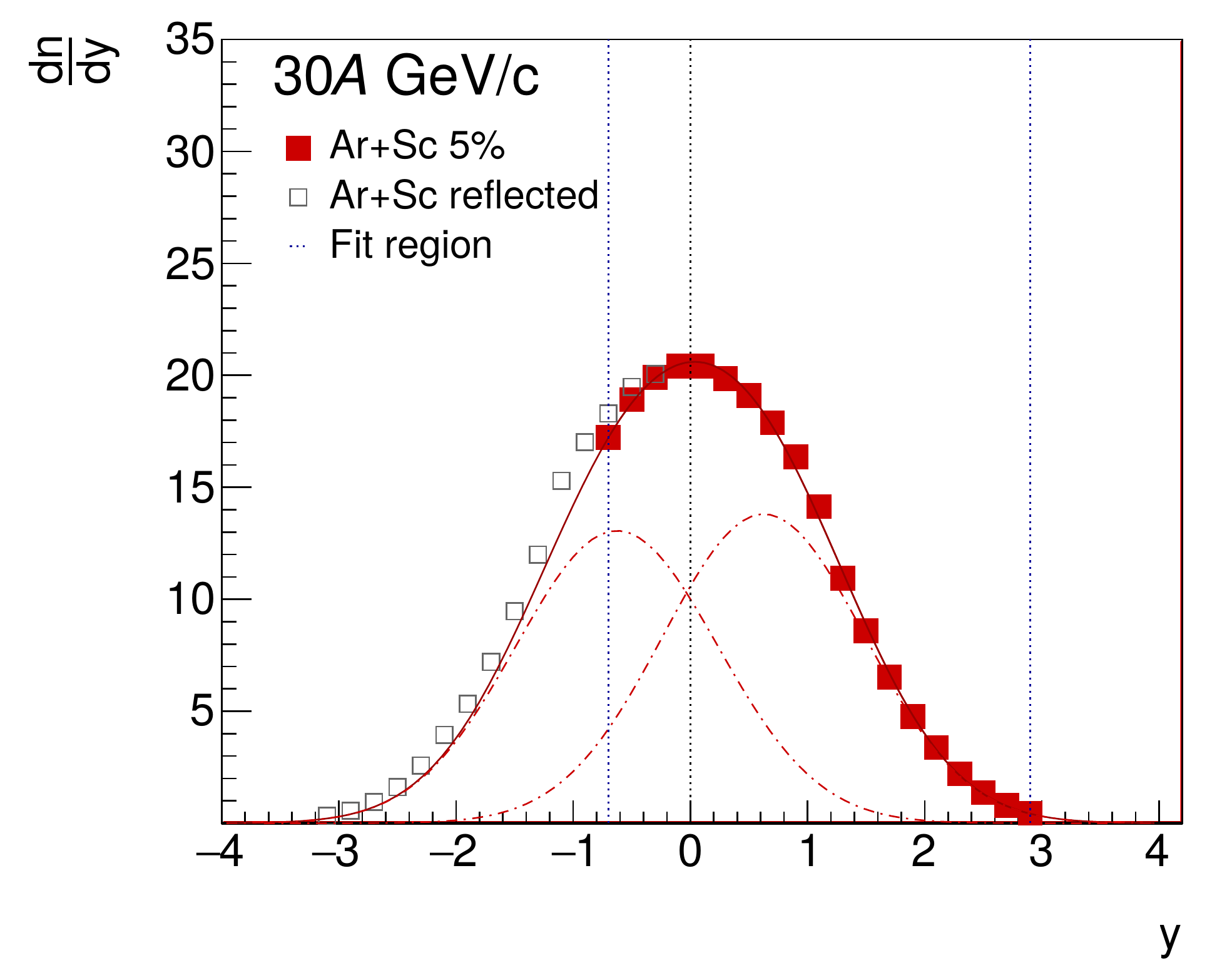}\\
  \includegraphics[width=0.3\textwidth]{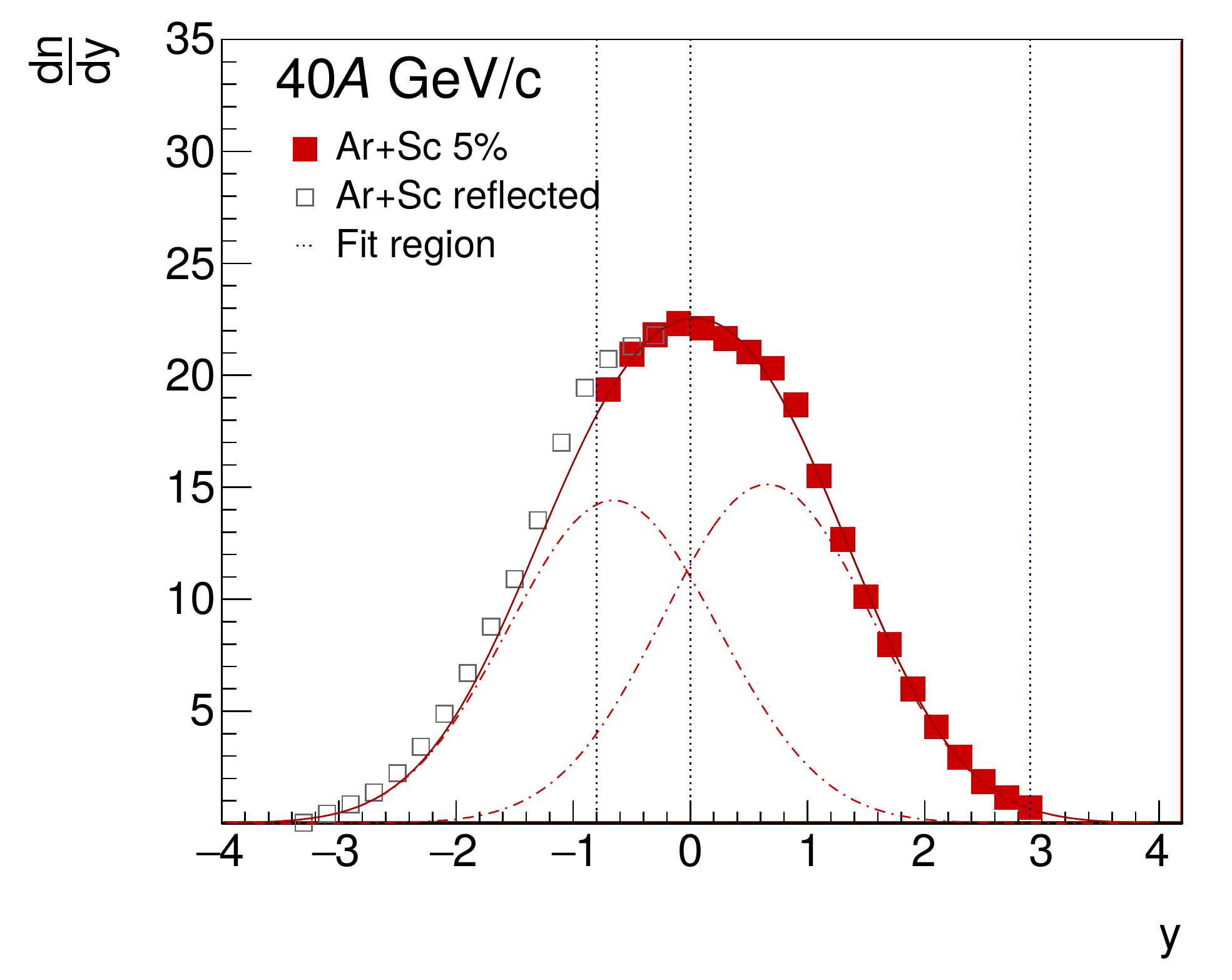}
  \includegraphics[width=0.3\textwidth]{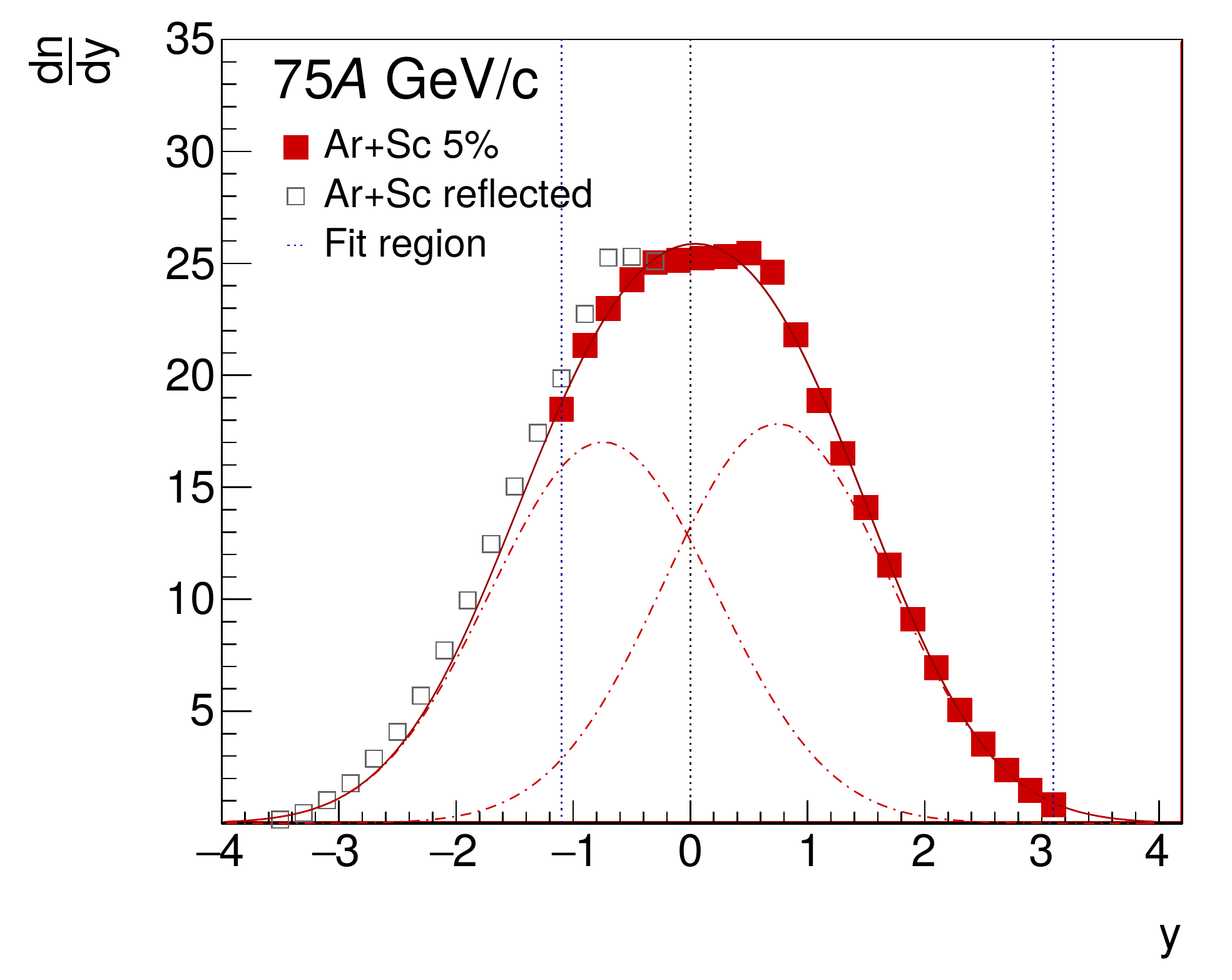}
  \includegraphics[width=0.3\textwidth]{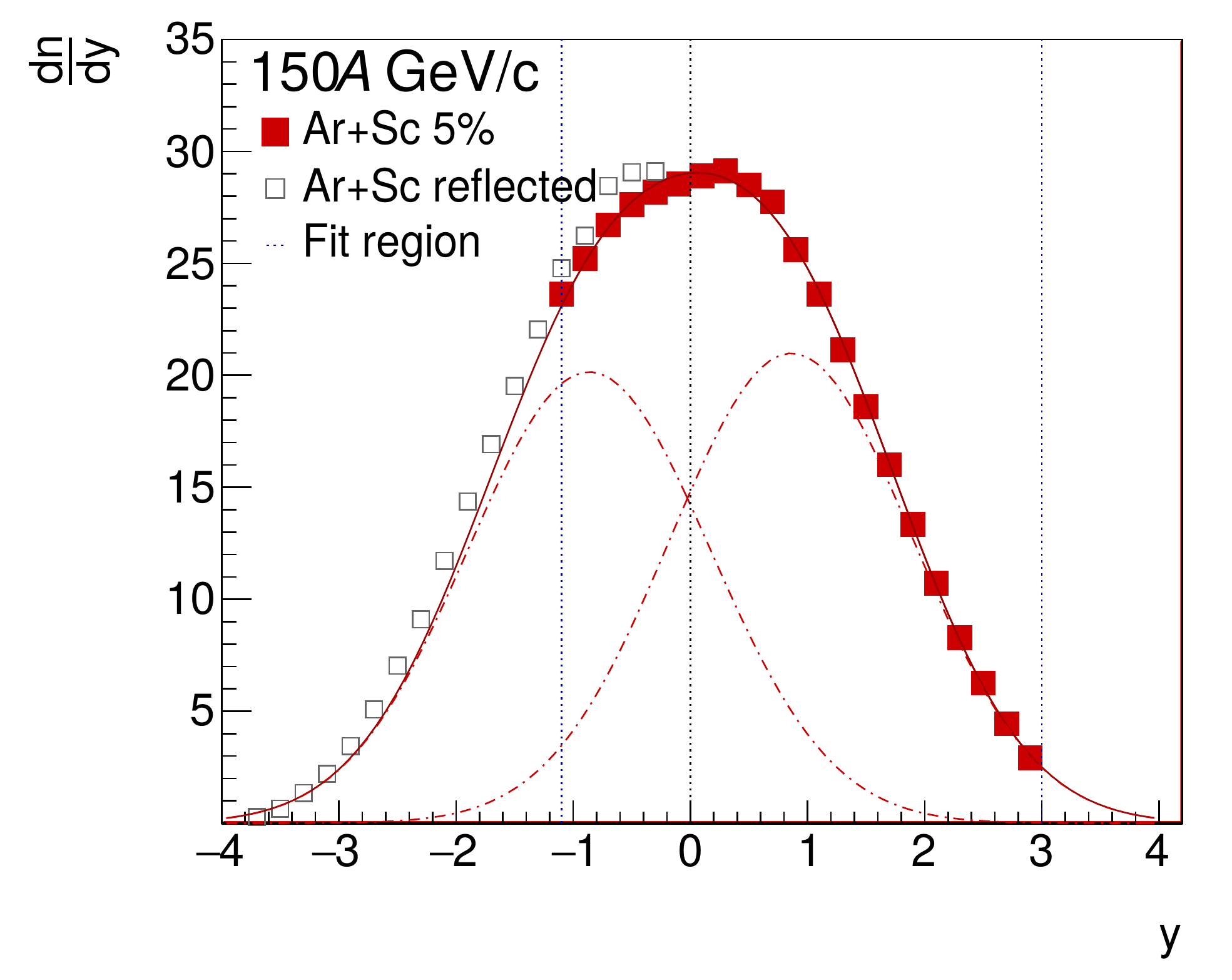}
  \caption{$\pi^-$ Rapidity distributions with fitted Gaussian functions for all measured beam momenta. The red boxes are measured points, open boxes are measured points reflected at midrapidity, dashed curves are fitted Gaussians and the solid curve is the sum of the fitted Gaussians.}
  \label{fig:rapidityDistr}
\end{figure}

The starting point for the analysis described herein are the $\pi^-$ rapidity distributions ${dn}/{dy}$ for the 5 \% most central collisions as shown in Fig.~\ref{fig:rapidityDistr}. They were obtained by M.~Lewicki using the so-called $h^-$ method wich is described in a separate contribution. Centrality was determined by selecting the 5 \% of collisions with the smallest forward going energy as measured by the projectile spectator detector.

Due to the wide acceptance of the NA61/SHINE experiment spectra could be extrapolated reliably to $4\pi$ acceptance. The extrapolation process consisted of two steps - extrapolation in transverse momentum $p_{\text{T}}$ for each bin of rapidity $y$ and extrapolation of ${dn}/{dy}$ in rapidity. For the latter a sum of two Gaussian functions was fiited, $g(y)=g_{\text{T}}(y)+g_{\text{P}}(y)$, where 

$$g_{\text{T}}(y)=\frac{A_0A_{rel}}{\sigma_0\sqrt{2\pi}}\exp\left(-\frac{(y-y_0)^2}{2\sigma_0^2}\right), g_{\text{P}}(y)=\frac{A_0}{\sigma_0\sqrt{2\pi}}\exp\left(-\frac{(y+y_0)^2}{2\sigma_0^2}\right)$$

Plots of the fitted functions together with measured data for all beam momenta are presented in Fig.~\ref{fig:rapidityDistr}.

\section{Mean multiplicities of $\pi^-$}

In order to calculate the mean negatively charged pion multiplicity $\langle \pi^- \rangle$, we utilized the following formula:

$$\langle \pi^- \rangle = \int_{-4}^{y_{\text{min}}}g(y)dy + \sum_{y_{\text{min}}}^{y_{\text{max}}} dy\left(\frac{dn}{dy}\right)_{\text{extrapolated in} p_{\text{T}}}+\int_{y_{\text{max}}}^4 g(y)dy$$

Thus the final result is the sum over measured values of ${dn}/{dy}$ in the acceptance region and the integral over the Gaussian fits outside. This procedure gave the results presented in table~\ref{tab:piMultiplicity}. Statistical uncertainties $\sigma_{\text{stat}}(\langle \pi^{-} \rangle)$ were obtained by propagating the statistical uncertainties of $\frac{dn}{dydp_{\text{T}}}$ spectra. Systematic uncertainties $\sigma_{\text{sys}}(\langle \pi^{-} \rangle)$ are assumed to be $5\%$ based on the previous NA61 analysis of p+p collisions.

\begin{table}
 \centering
 \footnotesize
 \begin{tabular}{l|cccccc}
  Momentum [\textit{A} GeV/c] & 13 & 19 & 30 & 40 & 75 & 150\\
  \hline
  $\langle\pi^-\rangle$ & $38.46$ & $48.03$ & $59.72$ & $66.28$ & $86.12$ & $108.92$\\
  $\sigma_{\text{stat}}(\langle \pi^{-} \rangle)$ & $\pm 0.021$ & $\pm 0.021$ & $\pm 0.024$ & $\pm 0.018$ & $\pm 0.0079$ & $\pm 0.0088$\\
  $\sigma_{\text{sys}}(\langle \pi^{-} \rangle)$& $\pm 1.92$ & $\pm 2.40$ & $\pm 2.98$ & $\pm 3.31$ & $\pm 4.30$ & $\pm 5.44$
 \end{tabular}
 \caption{Mean $\pi^-$ multiplicities in the 5 \% most central Ar+Sc collisions with systematic and statistical uncertainties.}
 \label{tab:piMultiplicity}
\end{table}

\section{Obtaining the number of wounded nucleons $\langle W\rangle$}

The estimate of the mean number of wounded nucleons was performed using two Monte Carlo models - Glissando 2.73~\cite{glauber} and EPOS 1.99 (version CRMC 1.5.3)~\cite{EPOS}. Uncertainties of $\langle W\rangle$ were not calculated and will not be presented. In both models event selection was based on the number of projectile spectators: collisions with the 5\% lowest number were chosen.

The differences between the resulting values of $\langle W\rangle$ are significant, see table~\ref{tab:wounded}. For lower momenta EPOS gives much smaller numbers than Glissando.

\begin{table}
 \centering
 \footnotesize
 \begin{tabular}{l|cccccc}
  Momentum [\textit{A} GeV/c] & 13 & 19 & 30 & 40 & 75 & 150\\
  \hline
  $\langle W\rangle_{\text{EPOS}}$ & 50.63 & 54.68 & 58.44 & 59.01 & 61.12 & 63.04\\
  $\langle W\rangle_{\text{Glissando}}$ & 67.44 & 68.85 & 68.98 & 69.01 & 68.87 & 69.18
 \end{tabular}
 \caption{Number of wounded nucleons $\langle W\rangle$ in central Ar+Sc collisions calculated for the 5~\% of collisions with the smallest number of projectile spectators using the Glissando and EPOS models.}
 \label{tab:wounded}
\end{table}

In order to stay consistent with previously used values, the result from the Glauber model based Glissando calculation $\langle W\rangle_{\text{Glissando}}$ was chosen. The resulting ratios of $\langle \pi^- \rangle/\langle W\rangle$ are presented in table~\ref{tab:ratios}.

\begin{table}
 \centering
 \footnotesize
 \begin{tabular}{l|cccccc}
  Momentum [\textit{A} GeV/c] & 13 & 19 & 30 & 40 & 75 & 150\\
  \hline
  $\langle\pi^-\rangle$ & $38.46$ & $48.03$ & $59.72$ & $66.28$ & $86.12$ & $108.92$\\
  $\langle W \rangle_{\text{Glissando}}$ & $67.44$ & $68.85$ & $68.98$ & $69.01$ & $68.87$ & $69.18$\\
  $\langle\pi^-\rangle/\langle W\rangle$ & $0.57$ & $0.69$ & $0.86$ & $0.96$ & $1.25$ & $1.57$\\
  $\sigma(\langle\pi^-\rangle/\langle W\rangle)$ & $\pm 0.028$ & $\pm 0.034$ & $\pm 0.043$ & $\pm 0.047$ & $\pm 0.062$ & $\pm 0.078$\\
 \end{tabular}
 \caption{Ratio of $\langle\pi^-\rangle/\langle W\rangle$ for the 5~\% most central Ar+Sc collisions using $\langle W\rangle$ estimated from the Glissando model.}
 \label{tab:ratios}

\end{table}

\section{$\langle\pi^-\rangle/\langle W\rangle$ and the 'kink' plot}

\begin{figure}
\centering
 \includegraphics[width=0.29\textwidth]{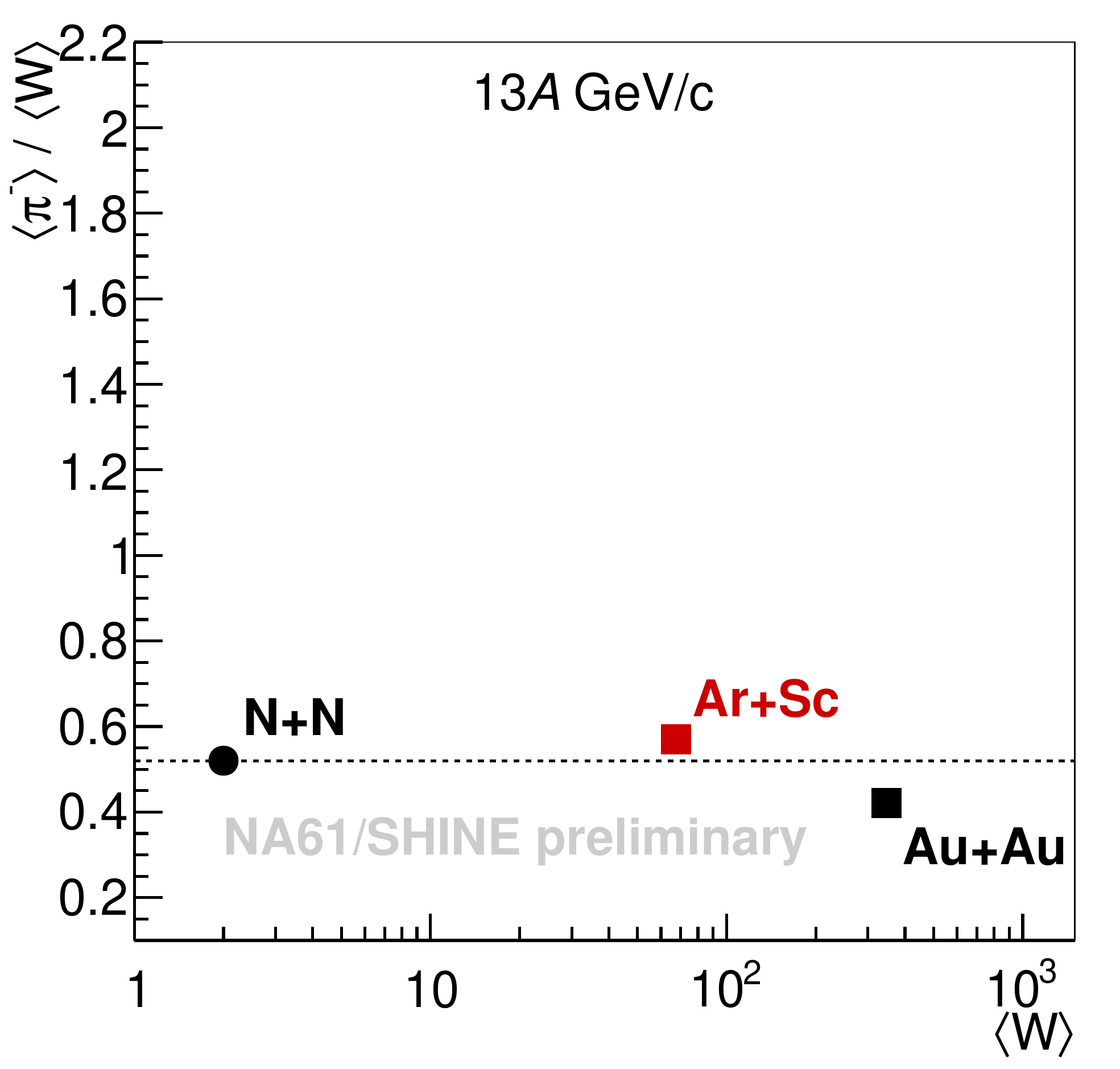}
 \includegraphics[width=0.29\textwidth]{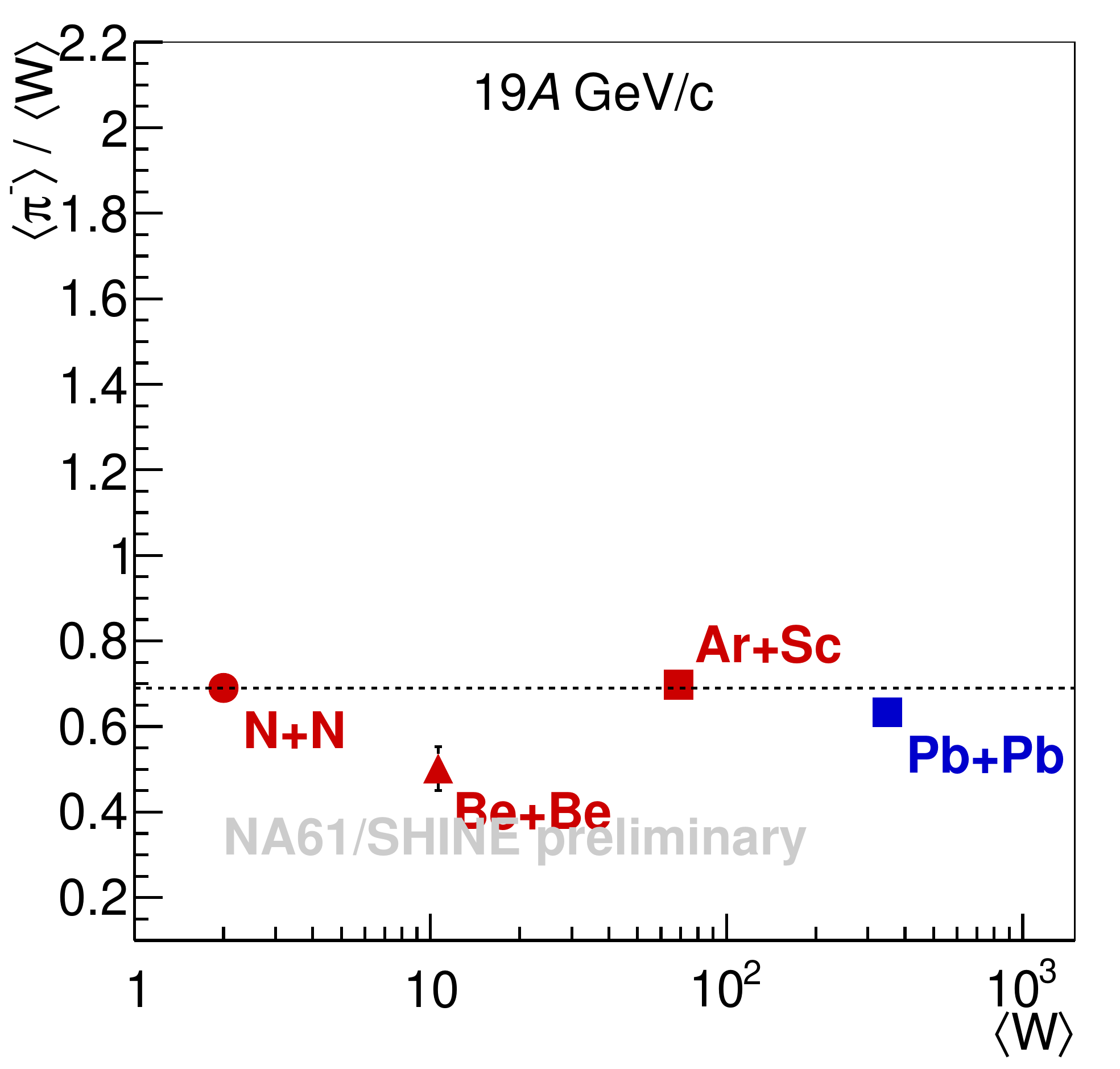}
 \includegraphics[width=0.29\textwidth]{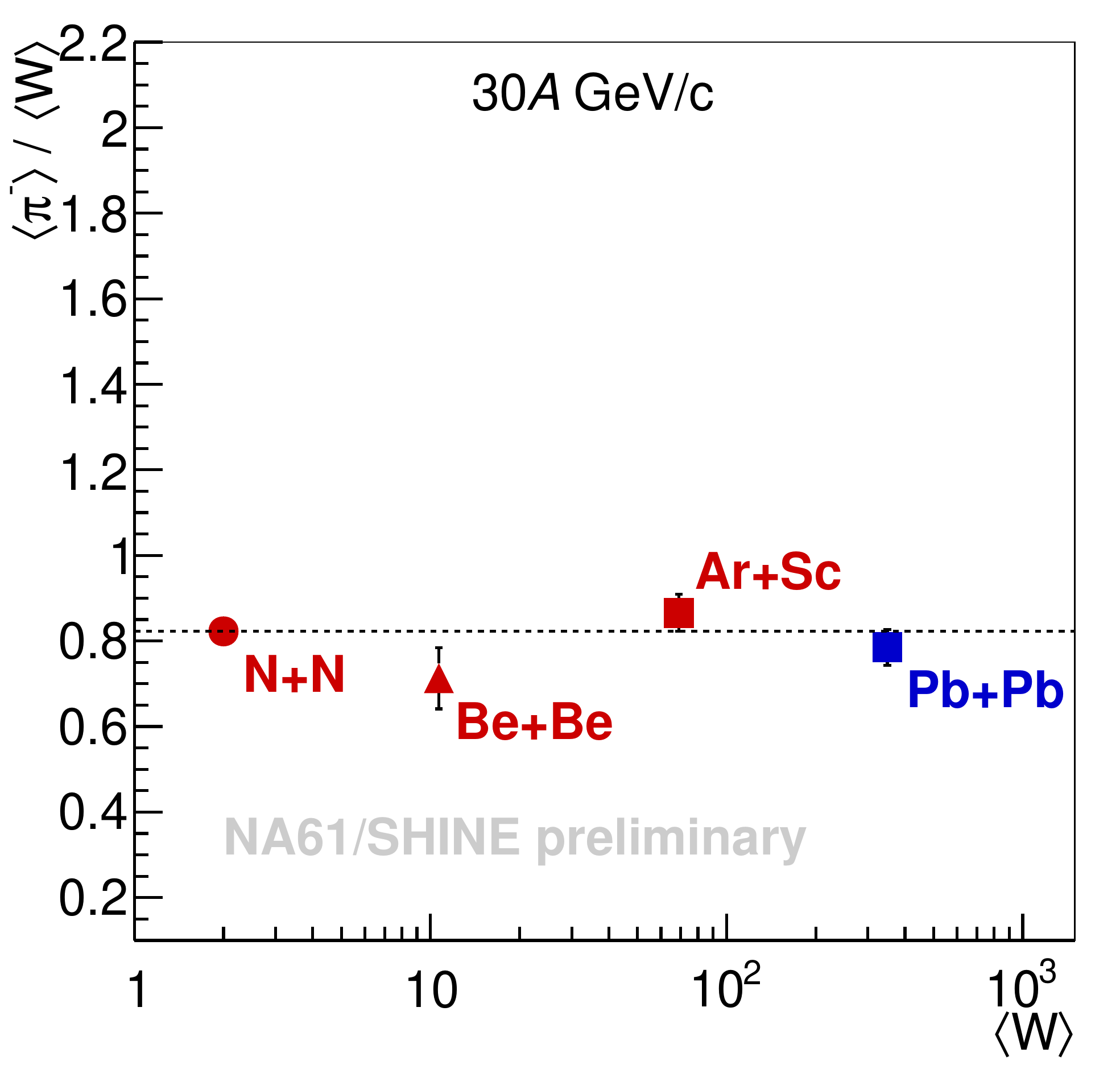}\\
 \includegraphics[width=0.29\textwidth]{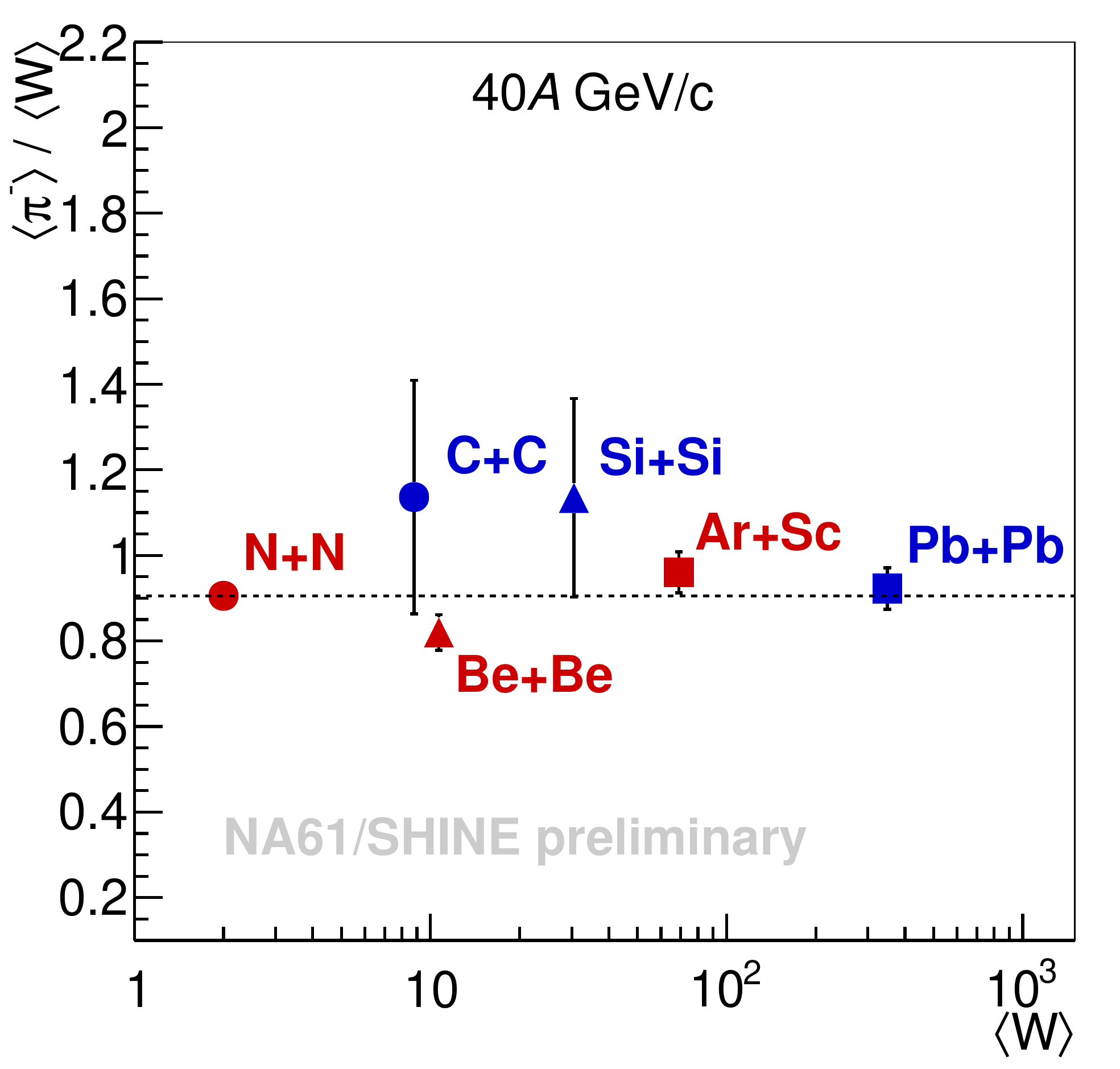}
 \includegraphics[width=0.29\textwidth]{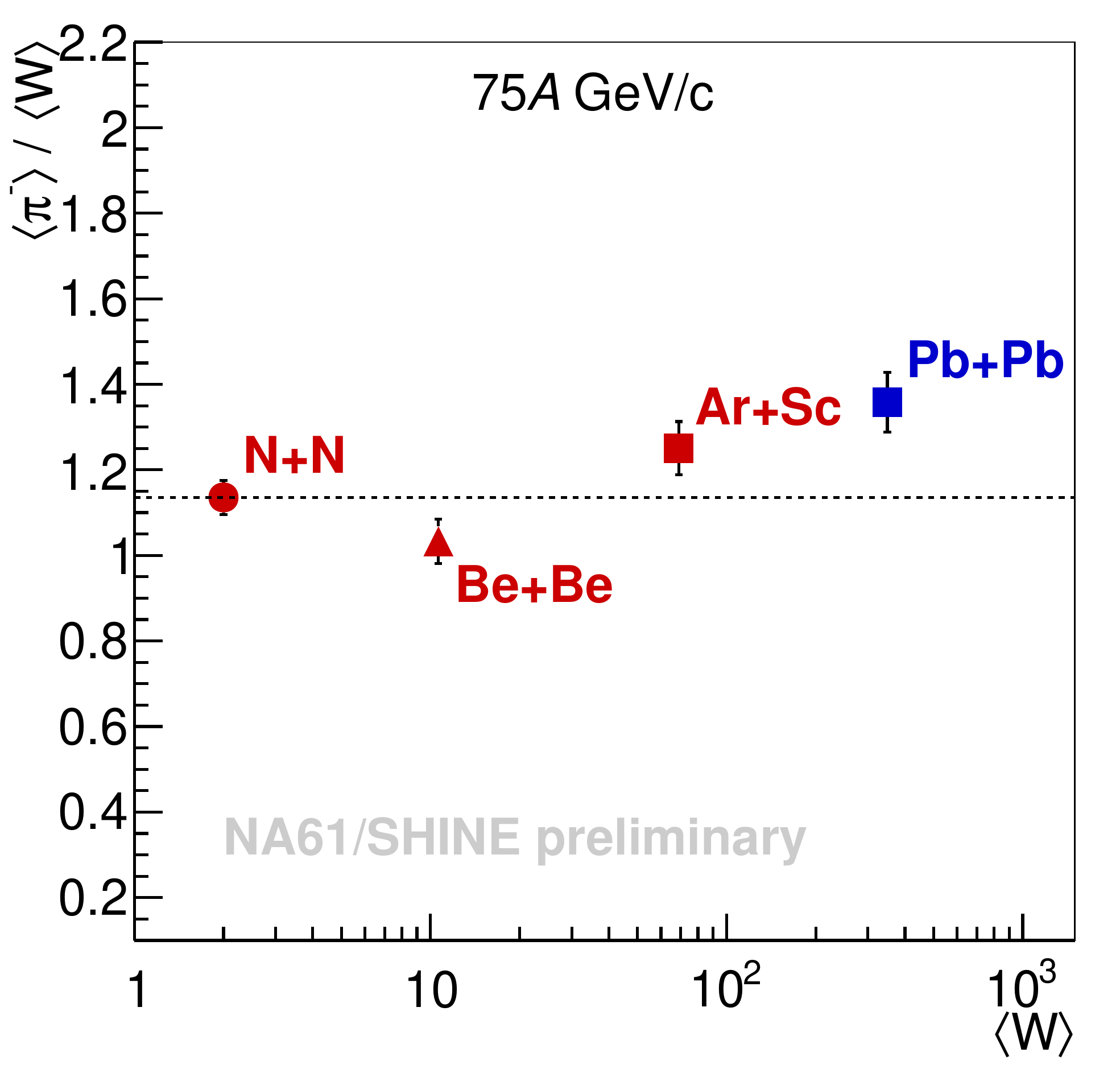}
 \includegraphics[width=0.29\textwidth]{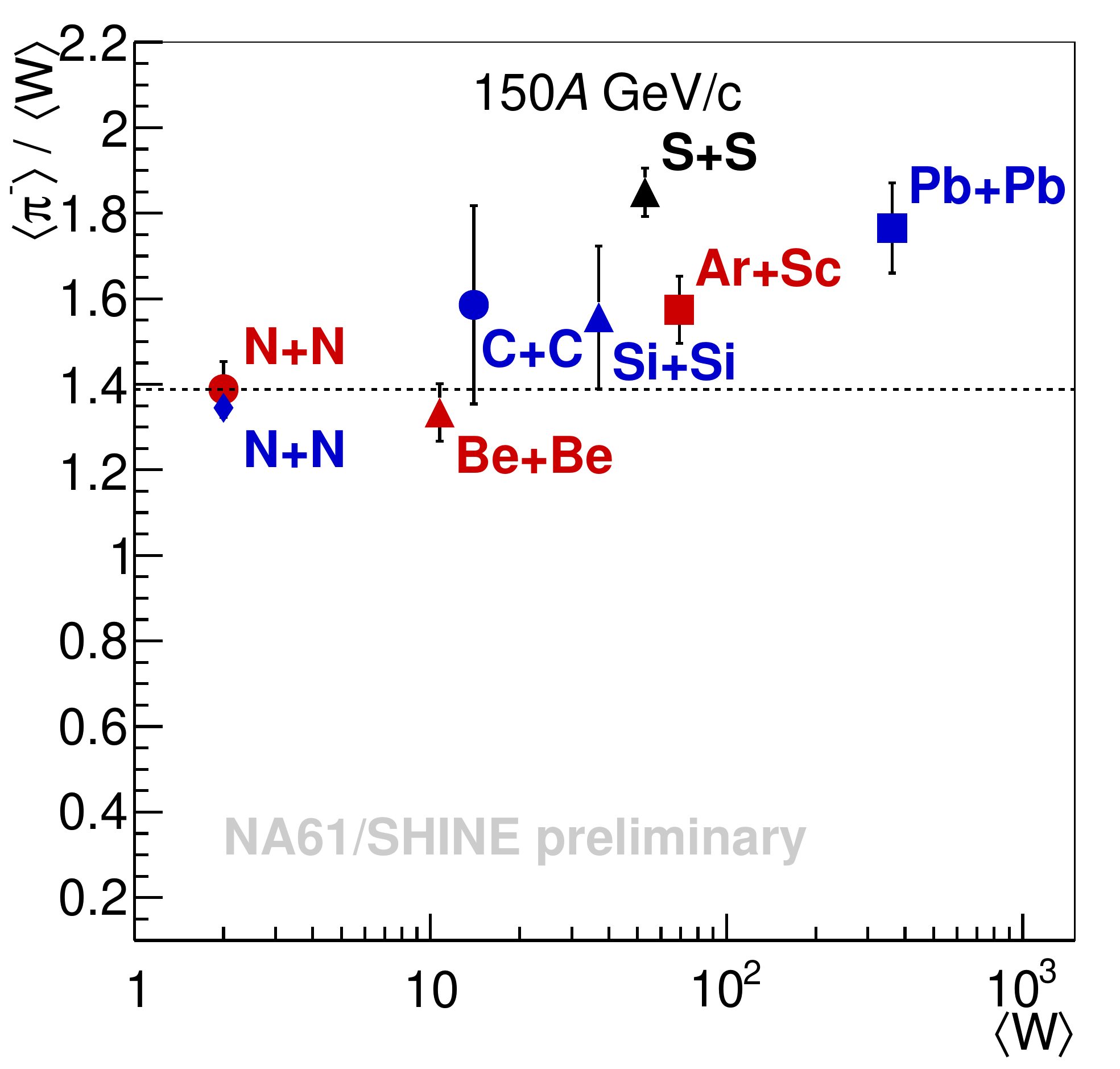}
 \caption{Comparison of the $\langle\pi^-\rangle/\langle W\rangle$ ratio from measurements in the SPS energy range.}
 \label{fig:comparison}
\end{figure}

The preliminary results on $\langle\pi^-\rangle/\langle W\rangle$ from central Ar+Sc reactions are plotted in Fig.~\ref{fig:comparison} together with other measuremets of NA61/SHINE~\cite{kaptur,antoni}, and data from NA49~\cite{PbPbLow,PbPbHigh,CCSiSiLow,CCSiSiHigh,ppNA49} and other experiments~\cite{AuAu,SS,ppBHM}. In order to compare results obtained for different systems, an isospin correction should be applied. To this end phenomenological formulas are used: $\langle\pi^-\rangle_{\text{N+N}}=\langle\pi^-\rangle_{\text{p+p}}+1/3$ and in the case of Au+Au the average $(\langle\pi^-\rangle_{\text{Au+Au}}+\langle\pi^+\rangle_{\text{Au+Au}})/2$. This procedure is based on the compilation of the world data and the model presented in \cite{scaling}. The correction for nucleus-nucleus collisions was only applied at 13$A$ GeV where it is large. As seen from Fig.~\ref{fig:comparison} the $\langle\pi^-\rangle/\langle W\rangle$ ratio shows little dependence on the system size at the lower energies while a rising tendency is observed at the highest energy. 

The $\langle\pi\rangle/\langle W\rangle$ ratio has often been plotted against the Fermi energy measure $F=\left[\frac{(\sqrt{s_{\text{NN}}}-2m_{\text{N}})^3}{\sqrt{s_{\text{NN}}}}\right]^{1/4}$.

\begin{figure}
\centering
 \includegraphics[width=0.7\textwidth]{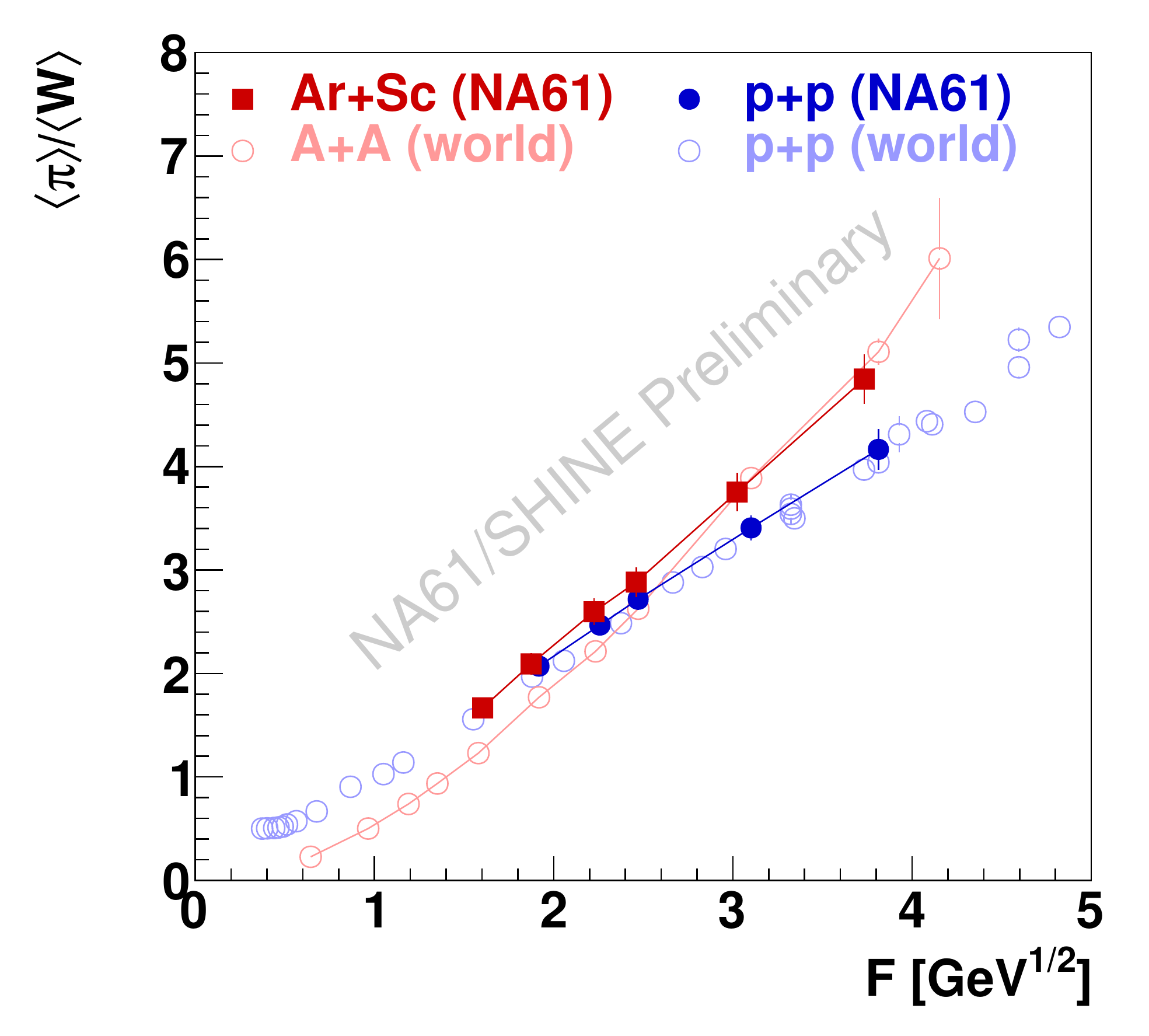}
 \caption{The "kink" plot with the preliminary Ar+Sc results.}
 \label{fig:kink}
\end{figure}

As for NA61/SHINE there are results only for $\langle\pi^-\rangle$ in Ar+Sc, Be+Be and p+p collisions, the multiplicities of $\langle\pi^+\rangle$ and $\langle\pi^0\rangle$ were approximated by $\langle\pi\rangle_{\text{p+p}}=3\langle\pi^-\rangle_{\text{p+p}}+1$ and $\langle\pi\rangle_{\text{Ar+Sc}}=3\langle\pi^-\rangle_{\text{Ar+Sc}}$. Figure~\ref{fig:kink} shows a compilation of world data compared to the NA61/SHINE measurements. One observes that for high SPS energies Ar+Sc follows the Pb+Pb trend, i.e. pion enhancement in central collisions of nuclei. On the other hand, for low SPS energies Ar+Sc follows the results from p+p reactions.

\FloatBarrier
\bibliographystyle{unsrt}
\bibliography{bibl.bib}

\end{document}